\documentclass[11pt,twoside]{article}
\usepackage{ap-article}
\usepackage{graphicx}
\usepackage{tikz}
\usepackage{caption}
\usepackage{pgfplots}
\usepackage{pgf-pie}
\usepackage[square,numbers]{natbib}
\usepackage{array}
\usepackage{blindtext}
\usetikzlibrary{patterns}
\pgfplotsset{compat=1.16}

\usepackage{afterpage}

\def\labart{yourLabel}      
\shortauthors{S. Jayasundara et al.}
\shorttitle{plant microRNA prediction}
\title{%
Machine learning for plant microRNA prediction: A systematic review
}
\author{%
Shyaman Jayasundara$^{1}$,~
Sandali Lokuge$^{1}$,~
Puwasuru Ihalagedara$^{1}$,~
Damayanthi Herath$^{1}$\footnote{Corresponding author: damayanthiherath@eng.pdn.ac.lk}
}
\addresses{%
$^{1}$
Department of Computer Engineering\\
University of Peradeniya, Peradeniya 20400, Sri Lanka
\\ \vspace*{0.05truein}
}
\begin{document}
\label{\labart-FirstPage}

\maketitle

\abstracts{%
 MicroRNAs (miRNAs) are endogenous small non-coding RNAs that play an important role in post-transcriptional gene regulation. However, the experimental determination of miRNA sequence and structure is both expensive and time-consuming. Therefore, computational and machine learning-based approaches have been adopted to predict novel microRNAs. With the involvement of data science and machine learning in biology, multiple research studies have been conducted to find microRNAs with different computational methods and different miRNA features. Multiple approaches are discussed in detail considering the learning algorithm/s used, features considered, dataset/s used and the criteria used in evaluations. This systematic review focuses on the machine learning methods developed for miRNA identification in plants. This will help researchers to gain a detailed idea about past studies and identify novel paths that solve drawbacks occurred in past studies. Our findings highlight the need for plant-specific computational methods for miRNA identification.
}

\medskip
\keywords{bioinformatics, novel miRNA, machine learning, microRNA, plant, systematic review, prediction}

\vspace*{7pt}\textlineskip
\begin{multicols}{2}

\section{Introduction}

MicroRNAs (miRNAs) are a recently discovered set of small, 20-24nt long, non-coding Ribonucleic acid (RNA) molecules. They are endogenously generated in plants and animals as a set of essential and evolutionary ancient components of genetic regulation\citep{he2004micrornas}. miRNAs silence their target genes by binding to the target messenger RNAs (mRNAs). This systematic review focuses on the discovery of plant miRNAs. Plant genomes encode more than hundreds of miRNA genes which are existing as families. DNA-dependent RNA Polymerase II (Pol II) transcribes miRNA genes and folds into a stem-loop structure called primary miRNA transcript (pri-miRNA). This pri-miRNA holds a 5' cap and a 3' poly-A tail \citep{Xie_2005}. In pri-miRNA in plants, the stem-loop varies in length (from 60 nt to over 500 nt) and have more complex structures than the equivalent ${\sim}70$ nt long animal stem-loop \citep{Xie_2005}. This pri-miRNA is then processed to pre-miRNA by Dicer-like RNase III endonucleases (DCLs) \citep{Bologna_2013,Song_2010}. Different plant species have different numbers of DCL proteins (e.g., in Arabidopsis thaliana DCL1 is the processing enzyme) \citep{Xuemei2005,Junli2019}. The result of the DCL mediated process with pre-miRNA is the miRNA/miRNA* duplex. However, plant miRNA biogenesis has an additional step called miRNA methylation done by HEN1 protein. Then, the methylated duplexes are exported to the cytoplasm by exportin 5, HASTY (HST) \citep{Xuemei2005,Junli2019}. One strand (miRNA) of the miRNA/miRNA* duplex is selected to join with Argonaute (AGO) protein. This selected strand is called the guide strand, and the complex is called the RNA-Induced Silencing Complex (RISC). For a long time, it was not clear where the RISC is assembled until confirmed that RISC is mainly assembled within the nucleus and exported to the cytosol by EXPO1 \citep{Bologna_2018}. miRNA*, called the passenger strand is discarded. The guide strand is the one responsible for the gene silencing.

miRNAs are involved in various biological processes in plants such as development and growth, genome integrity maintenance, innate immunity, hormone signaling pathways, and response to different environmental abiotic and biotic stresses \citep{Xie_2014, Zhang_2015}. For instance, miR156, miR172, and miR396 in Cotton (\emph{Gossypium hirsutum}) are responsible for stress-responsiveness, flower development, and Cotton fiber development, respectively. Some miRNAs are found in multiple plant species (for example, miR156 is found in Tea plant (\emph{Camellia sinensis}), Cotton, and Tobacco (\emph{Nicotiana tabacum})). miRNA-based genetic modification technology is widely used in agriculture as they are directly related to biological and metabolic processes in economically important crops. For instance, down-regulation of miR169, up-regulation of miR389, and overexpression of osa-miR396c are observed in drought-tolerant, thermotolerant, and salt and alkaline tolerant varieties respectively \citep{Peng2010,Trindade2010,Ramanjulu2012}. Thus, artificial miRNAs equivalent to specific miRNAs are developed to change the biological processes of the plants forcibly to enhance their productivity and cure various plant diseases.

At present, there are four methods for identifying miRNAs \citep{Zhang2006}: genetic screening \citep{wightman_ha_ruvkun_1993}, direct cloning after isolating small RNAs \citep{lu_2005}, computational strategies\citep{Brown2005}, and Expressed Sequence Tags (ESTs) analysis \citep{Zhang2005}. In the beginning, genetic screening was used for miRNAs discovery. However, this method was time-consuming, expensive, and subjected by chance. In the next experimental approach, small RNA molecules are isolated using size fractioning. After that, the small RNAs are ligated with RNA adapters and reversed transcribed into cDNAs. This method overcomes the weaknesses of the genetic screening method to some extent since the procedure isolates and screens only the small RNAs. However, miRNAs expressions at lower levels, RNA degradation during sample separation, and physical properties such as sequence composition can make the cloning process difficult \citep{prabu2010computational}. EST analysis can be used to identify evolutionarily conserved miRNAs in related species. This approach is more suitable to recognize conserved miRNA in closely related species whose genomes are unknown or are poorly understood. Though we can predict homologs and orthologs of known miRNAs, most miRNAs are not evolutionarily conserv-ed. Therefore, the possibility of finding new miRNAs throu-gh this method is low.

Studying microRNAs is challenging because microRNAs are very short, sometimes the miRNA sequences of closely related microRNA families differ only one nucleotide. Therefore, the new approaches should possess high specificity and should be able to recognize even a single-nucleotide mismatch. Aside from the above-mentioned laboratory experimental methods, computational methods are efficient, less expensive, and time-saving alternatives for the novel miRNAs discovery in plants and animals. The underlying principle behind these computational methods is to learn from examples (i.e., known miRNAs) and predict novel miRNAs.

There are mainly three types of bioinformatics approaches for novel miRNA discovery; comparative and homology-based approaches, ab-initio or non-comparative approaches, and integrated approaches \citep{Koh2017}. In comparative and homology-based approaches, phylogenetic conservation of primary or/and secondary structures of the pre-miRNAs of known miRNAs are used to predict novel miRNAs similar to them. This approach is effective in predicting miRNAs when the sequences of considered species are closely conserved. But they may not be effective enough to predict miRNAs of divergent sequences. Ab-initio methods use computational models that can identity novel miRNA in a broader range of unseen sequences. Instead of doing a direct comparison, ab-initio methods are trying to identify the inherent structural or compositional features of miRNAs. Machine learning (ML) is a discipline that uses statistical methods to learn from examples and makes predictions on unseen data based on the gained experience \citep{alpaydin2020introduction}. ML algorithms are mainly categorized under supervised ML algorithms and unsupervised ML algorithms. Supervised ML algorithms take advantage of known labeled data to learn from, and make predictions on future data. In contrast, unsupervised ML algorithms use unlabeled data. Usually, supervised ML algorithms are used in ab-initio approaches \citep{Koh2017,Meng2014,Tseng2018,Jiang2007}. Increasing the specificity of the prediction algorithm by reducing the number of false-positives miRNA prediction is the most challenging part of the ML models.

This review focuses on the studies which have incorporated ML methods for microRNA identification in plants. It discusses the datasets used in the focused studies, features of miRNAs considered in different studies, algorithms used in ML models, model evaluation, and the performance of the models.

\section{Study Selection}

This section describes how we selected the publications on the topic and focused areas from the previous studies. This review focuses on the literature published in the last ten years. We chose publications from the databases of Google Scholar, Science Direct, and PubMed (as of 10 Feb 2020).  We used the following term to search the publications related to the topic from the mentioned databases. 

\begin{itemize}
  \item plant AND (miRNA OR microRNA) AND (classification OR prediction OR identification OR discovery) AND (``machine learning'')
\end{itemize}

The search results of Google Scholar, Science Direct, and PubMed were 4490, 137, and 39, respectively. To narrow-down the 4490 Google Scholar results, we only considered the output of the first ten pages of the search results. The results from  Science Direct and PubMed were directly taken without narrowing-down.

Out of the search results from the three databases, we filtered the articles by evaluating the titles with our exclusion and inclusion criteria (Table \ref{tab:selec_cri}). Secondly, the abstracts were also considered for some studies when it was difficult to judge by the title. Further, we referred to the contents of the publications for the filtration concerning our exclusion and inclusion criteria. After following these three steps, the number of publications selected from Google Scholar, Science Direct, and PubMed is 22, 3, and 12, respectively. 

\medskip

\begin{Table}\centering 
    \tcap{Article Selection Criteria}
    {
    \small
    \setlength{\extrarowheight}{3pt}
    \baselineskip=13pt
    \begin{tabular}{ p{3.2cm}|p{3.8cm} }
    \hline
        Inclusion criteria & Exclusion criteria  \\
    \hline
        Studies that use machine learning &  Studies that use different techniques (sequence homology, NGS techniques)\\
    \hline
        Studies that use the plant data & Studies that use only human, animal, virus data and plant data with the human, animal, worm, and vertebrate data \\
    \hline
         \multirow{3}{3.2cm}{Journal/conference proceeding publications outlining methods}  & Literature reviews/surveys on the subject  \\
    \cline{2-2}
         & Studies that have predicted miRNA target genes  \\
    \hline
    
    \end{tabular}
    }
    \label{tab:selec_cri}
\end{Table}

\medskip

\begin{Figure}
\centering
\includegraphics[width=3.1in]{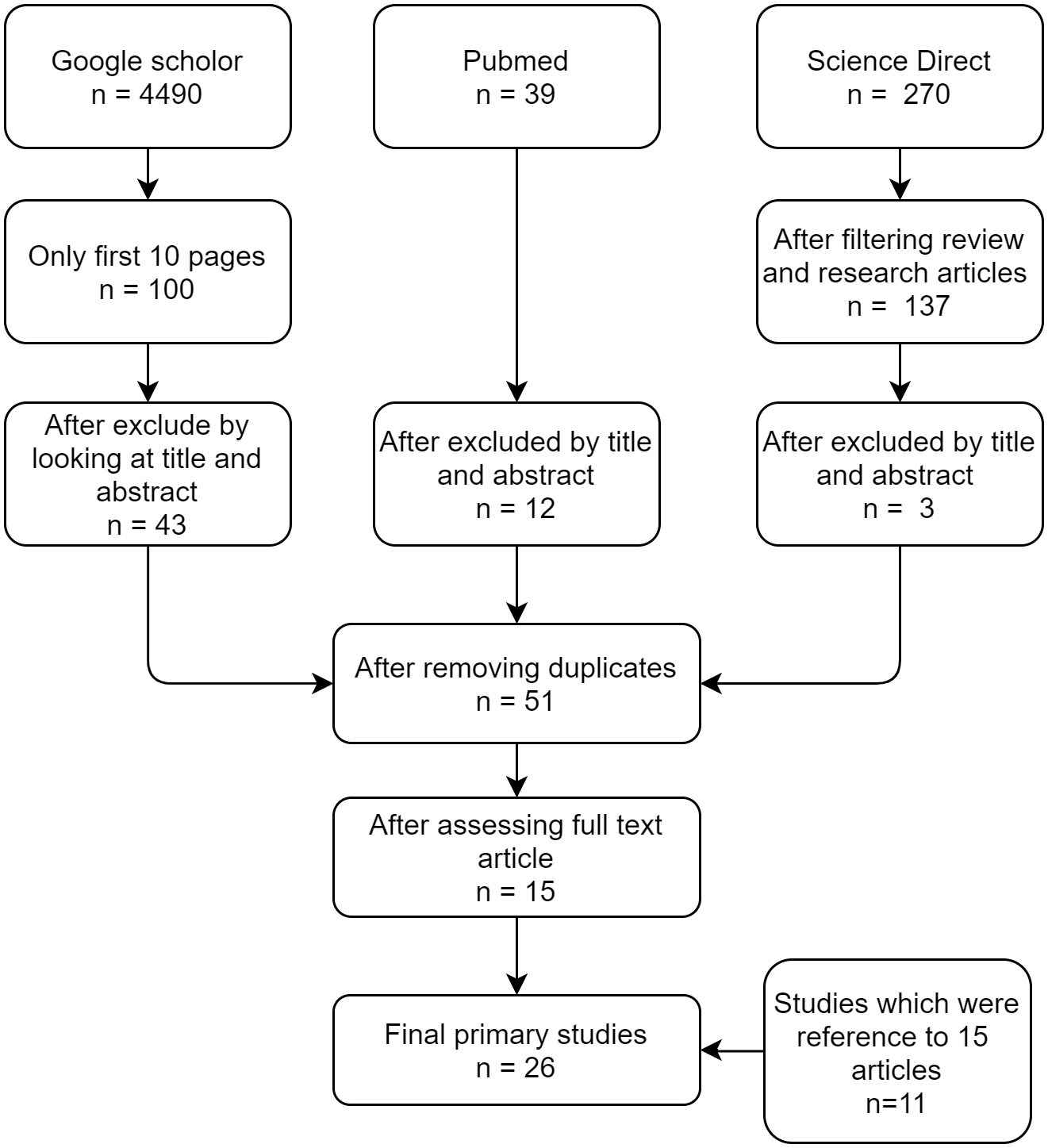}
\captionsetup{margin=10pt,font=small,labelfont=bf, justification=centering}
\fcaption{Flowchart of article selection}    
\label{fig:paper_exclusion_flowchart}
\end{Figure}

After eliminating the duplicates, a total number of 15 studies remained for review. Further, we included 11 more papers besides the search strategy due to the high number of citations. These additional papers include some studies conducted on animal pre-miRNAs (humans, viruses). However, the studies on plant pre-miRNA identification have cited those studies because of their significant impact on this study area. Therefore, we added them to our review considering their contribution to the novel miRNA discovery. Hence, our systematic review was conducted including 26 studies (Fig \ref{fig:paper_exclusion_flowchart}).

\section{Datasets used for model training and testing}

This section describes the nature of the datasets used to train and test the models in the considered studies. 

The selected studies related to this topic can be divided into two groups by considering the dataset as only plant species datasets and plant and animal species combined datasets. The studies \cite{Meng2014}, \cite{Tseng2018}, \cite{Douglass2016}, \cite{Williams2012} have used datasets that consists with only plant species. Some studies like \cite{Koh2017}, \cite{Vitsios2017}, \cite{Zhong2015}, \cite{Wu2011} have used both animal and plant datasets in their studies. The study \cite{Xiao2011} has trained its model with pre-miRNAs of animal species and tested with pre-miRNAs of plant species and viruses.

Some of the studies discovering plant miRNAs have used datasets of multiple species. For instance, \cite{Meng2014} and plantMirP\cite{Yao2016} use 9 plant species and  \cite{Williams2012}, \cite{Douglass2016} use 18 and 4 plant species respectively. Similarly, some studies \cite{Tseng2018}, \cite{Sunkar2008} have only focused on specific species like Arabidopsis, Soybean, Orchids, and Rice.

Since most of the studies are based on supervised machine learning models, they have separated the data set into two classes named positive and negative. Then these labeled data are used to train the machine learning model. 

Almost all the studies use a version of an online repository for the RNA sequences and annotations, namely the miRBase database as a source of positive dataset. The most frequently used plant species for the positive dataset are \emph{Glycine max, Arabidopsis thaliana, Oryza sativa, Zea mays, Sorghum bicolor, Arabidopsis lyrata, Physcomitrella patens, Populus trichocarpa} and \emph{Medicago truncatula}. Some studies have used only experimentally validated miRNAs as a positive dataset \citep{Meng2014,Douglass2016}. Similarly, studies have used other criteria such as both reads of a read pair should be mapped to the same miRNA precursor \citep{Williams2012}, the read count number of one strand should be bigger than 10 and the read length has to be between 20 to 24 nt \citep{Tseng2018} to construct the positive dataset. PlantMirP is another study that has collected all miRNAs of \emph{Viridiplantae} after removing all sequences that contain non-AGCU characters. The study \cite{Yousef2015a} has used all available miRNAs for selected eight plant species from miRBase (release 20 and 21). Along with that, they have considered data published on the web server PlantMiRNAPred\citep{Xuan2011}. Studies like \cite{Zhong2015} uses datasets from previous studies.
 
The negative dataset consists of a range of data to ensure they are free from real miRNAs. Different methods and different data sources have been used to generate a quality negative dataset. The majority of the studies use coding regions (CDS) of human, animal, and plant species with different criteria to generate the negative dataset \citep{Tseng2018,Xuan2011}. These CDS are downloaded from different databases like Phytozome \citep{goodstein2012phytozome}, Ensembl Plants \citep{enseblplants}, and The Arabidopsis Information Resource (TAIR) \citep{tair}. The study \cite{Meng2014} randomly gathered parts of hairpin sequences in the coding region from four different plant species using a 60-150 nt sized window. After that, they used two criteria to select the negative dataset. A negative dataset is created with Expressed Sequence Tags (ESTs) downloaded from TIGR Plant Transcript Assemblies \citep{Williams2012}. The short sequences are randomly selected from the central region of ESTs and filtered by looking at Shannon entropy and other criteria. 

\begin{figure*}[!ht]
\begin{tikzpicture}
\begin{axis}[
    ybar, 
    ymin=0,
    xtick = data,
    height=70mm,
    width=\linewidth-2pt,
    xlabel = {Study},
    ylabel = {Number of sequences},
    x tick label style={rotate=45,anchor=east},
    xtick pos=left,
    ytick pos=left,
    symbolic x coords = 
    {
    \citet{Yousef2016_2},
    \citet{Xuan2011_1},
    \citet{Silla2010},
    \citet{Engchuan2012},
    \citet{Xue2005},
    \citet{Batuwita2009},
    \citet{Yousef2015},
    \citet{Xuan2011},
    \citet{Sunkar2008},
    \citet{Yao2016},
    \citet{Xiao2011},
    \citet{Jiang2007},
    \citet{Guan2011},
    \citet{Meng2014}
    }
  ]

\addplot[black, pattern = crosshatch dots]
coordinates { 
(\citet{Jiang2007},163) 
(\citet{Meng2014},2000)
(\citet{Yousef2015},980)
(\citet{Xiao2011},3000)
(\citet{Yao2016},3044)
(\citet{Xuan2011},980)
(\citet{Sunkar2008},242)
(\citet{Guan2011},527)
(\citet{Batuwita2009},691)
(\citet{Xue2005},163)
(\citet{Engchuan2012},872)
(\citet{Silla2010},430)
(\citet{Xuan2011_1},1323)
(\citet{Yousef2016_2},1332)
};
\addplot[fill=black!40]
coordinates { 
(\citet{Jiang2007},168) 
(\citet{Meng2014},2000)
(\citet{Yousef2015},980)
(\citet{Xiao2011},3000)
(\citet{Yao2016},5186)
(\citet{Xuan2011},980)
(\citet{Sunkar2008},242)
(\citet{Guan2011},1971)
(\citet{Batuwita2009},168)
(\citet{Xue2005},972)
(\citet{Engchuan2012},980)
(\citet{Silla2010},400)
(\citet{Xuan2011_1},1323)
(\citet{Yousef2016_2},980)
};

\legend{Negative,Positive}
\end{axis}
\end{tikzpicture}
\fcap{Number of real pre-miRNAs and pseudo pre-miRNAs in training sets}
\label{fig:pos_neg}
\end{figure*}

Certain differences in terms of the number of the positive and negative datasets used in studies can be observed. Figure \ref{fig:pos_neg} represents the number of pre-miRNA sequences and pseudo pre-miRNA sequences which have been used to train the ML models in plant pre-miRNA identification. According to Figure \ref{fig:pos_neg} most of the studies have used an equal amount of sequences for positive and negative training datasets. Similarly, there are studies which have used a high number of RNA sequences to the negative data set than the number of miRNAs in the positive dataset. That may affect the performance of the models.

\section{Machine Learning approaches}

This section covers the computational approaches used by the selected studies to develop plant miRNA, identification models. The computational methods for miRNA identification can be grouped as comparative and non-comparative methods. Both groups have their advantages and limitations. But the non-comparative methods which are based on machine learning algorithms have risen beyond the comparative methods. It is due to the inability of comparative methods to identify novel miRNAs that do not share sequence homology with known miRNAs. (In the comparative method, it searches the exact or near-exact sequence which matches to previously known miRNAs). 

In machine learning-based methods, the model learns the sequence and structural features of miRNAs.  Typically, the inputs to the machine learning models are a set of features that describes a candidate miRNA (have discussed under Section 5) and the output would be either 1 or 0 (-1) indicating whether the candidate is miRNA or not. 

Support Vector Machine (SVM), Naïve Bayes, Decision trees, Random forest, and AdaBoost algorithm have been used as machine learning algorithms in the selected studies.  Figure \ref{fig:ml_algo_pie} shows the distribution of the mentioned machine learning algorithms used in the selected studies. Majority of the considered studies have used the SVM algorithm for the model implementation \citep{Koh2017, Meng2014, Tseng2018} (Fig. \ref{fig:ml_algo_pie}). The study \cite{Meng2014} has trained two SVM classifiers to identify pre-miRNAs mature miRNAs. Those SVM models have displayed high overall performance in discovering novel MiRNAs in plants. Some of the studies have combined an adaptive boosting algorithm with SVM (AdaBoost-SVM) to transfer weaker classifiers into one stronger classifier \citep{Wang2019}. This combined algorithm has shown a high degree of efficacy compared with the other studies in the mature miRNAs identification. The study \cite{Guan2011} has developed a miRNA detection model based on the AdaBoost algorithm with a set of novel transition probability matrices (TPM) and novel miRNA biogenesis vectors. This is the first visual miRNA prediction algorithm that addresses the problems encountered in ab-initio miRNAs prediction models such as the inability to being applied to multi-species data, and the inability to balance between specificity and sensitivity.

Random Forest is the second most widely used algorithm. The Random Forest algorithm-based models have high accuracy and less computation time \citep{Jiang2007, Zhang2005}. The prediction model of the study \cite{Zhang2005} has reduced their variance and has avoided overfitting. The tree structure has given the advantage of measuring the importance of features. Study \cite{Vitsios2017} has followed the Decision Forest algorithm to construct the miRNA prediction model from small RNA sequencing data. Similarly, plant miRNA have been identified by utilizing Bayesian approaches. The study \cite{Douglass2016} is one of the research that has used the Naïve Bayes algorithm without eliminating any small RNAs from a smRNA-seq library. As shown in Figure \ref{fig:ml_algo_pie}, the Decision Tree algorithm has less involvement in the plant miRNA model construction. However, \cite{Williams2012} has shown that the C5.0 Decision Tree algorithm can be used to produce a highly accurate universal plant miRNA predictor. 

A limited number of studies used the combination of both comparative \& homology-based methods and the ab-initio method for miRNA prediction. MiRHunter \citep{Koh2017} is such an example. They have used a hybrid method using a comparative \& homology-based method and the ab-initio method. Initially, evolutionary conservation and hairpin structure as preliminary filters are used to select the preliminary pre-miRNA candidates by following a support vector machine classifier. In the aforementioned work, a large number of false-positives from the filters have been eliminated by using the machine learning model. Their cross-validation (CV) scores show that the hybrid method has performed well than using a machine learning model only.

Since the negative class cannot be established experimentally, most of these methods require the generation of an artificial negative class which may lead to problems \citep{Allmer2012}. Selecting positive examples (real miRNAs) is usually straightforward. But the negative data construction is harder since there is an uncertainty in deciding whether a candidate is a pre-miRNA. Therefore, the lack of quality negative class datasets is a serious issue in building machine learning models to predict miRNAs. The study \cite{Yousef2015} has focused on an ab-initio method of one-class classifiers to identify miRNAs as a remedy.

\begin{Figure}
\centering
\begin{tikzpicture}[scale = 1]
\pie[color={black!45, black!35, black!25, black!15,black!5},rotate =215,radius=2.3]{47.8/SVM,21.7/Random Forest,17.4/Naive Bayes, 8.7/AdaBoost,4.3/Decision Tree}
\end{tikzpicture}
\fcaption{Machine learning algorithms used}
\label{fig:ml_algo_pie}
\end{Figure}

\section{Features considered in prediction models}
To identify and discover microRNAs in plants, different types of unique features of miRNAs have been used over the years. This section describes the features used to develop machine learning models. 

The principles of the machine learning approach for miRNA or miRNA precursor (pre-miRNA) discovery are based on three major characteristics (features) of miRNAs or pre-miRNAs; structural, thermodynamical, and sequence-based features \citep{buwani2019}. Features such as hairpin length, hairpin-loop length, bulge size and location, base-pairing, minimum free energy (mfe), triplet elements, and distance of the miRNA from the loop of the hairpin precursor are under structural features. Thermodynamical features include entropy and enthalpy measures of the structures. Nucleotide content and location, repeat elements, sequence complexity, motifs, and n-grams are under sequence-based features. In addition to these features, there are some evolutionary conservation features such as conserved motif and signature, sequence and structure similarity, and evolutionarily biased sequence composition from species to species \citep{Koh2017}. The secondary structures that are needed to compute features like mfe, structural diversity, thermodynamical features and base-pair related features are calculated by RNAfold \citep{rnafold} in ViennaRNA \citep{vienna2003} package \citep{Xuan2011,Tseng2018,Xue2005,Yao2016}.

Some of the studies have used the same features which were used in miPred\citep{Jiang2007}, microPred\citep{Batuwita2009}, Triplet-SVM\citep{Xue2005} and plantMiRNAPred approaches. Though miPred and microPred are based on human pre-miRNA, their features are used in plant pre-miRNA identifications \citep{Meng2014,Yao2016,Xuan2011}. In the study \cite{Meng2014}, 69 novel features are introduced besides the features obtained from past studies. They have combined thermodynamical features with \%(G+C) contents at the beginnings and endings of the sequences. For instance, MFE Index 7, MFEI7 = MFE/\%G + C\_ Begin\_n\_ 21nts, where \%G + C\_ Begin\_ n\_21nts is the GC content in the first 21 bases of the stems.

\begin{table*}
    \tcap{Number of features used in the selected studies}
    {
    \centering
    \small
    \begin{tabular}{|p{0.21\textwidth}|>{\centering\arraybackslash}p{0.14\textwidth}|>{\centering}p{0.14\textwidth\arraybackslash}|>{\centering\arraybackslash}p{0.13\textwidth}|>{\centering\arraybackslash}p{0.24\textwidth}|} 
    \hline
        Study & Thermodynamics features & Sequence-based Features &  Structure-based Features & Data\\
    \hline
        \citet{Koh2017} & 2 &17 &10& Plants, Animals, Viruses \\
        \citet{Meng2014} & 8 &17 &127& Plants \\
        \citet{Tseng2018} & - &1 &4& Plants \\
         \citet{Guan2011} & - &15 &9& Human, Animal, Plants, Viruses \\
        \citet{Jiang2007} & - &- &34& Human \\
        \citet{Douglass2016} & 1 &1 &3& Plants \\
        \citet{Williams2012} & 5 &14 &10& Plants \\
        \citet{Vitsios2017} & - &24 &9& Animals and Plants \\
        \citet{Zhong2015} & - &81 &58& Human, Plants, Animals \\
        \citet{Wu2011} & - &6 &6& Animal, Plants, Viruses (pre-miRNA)\\
        \citet{Xiao2011} & - &- &24& Animals, Plants, Viruses \\
        \citet{Yao2016} & - &17 &36& Plants \\
        \citet{Sunkar2008} & - & $<$20 &-& Plants \\
        \citet{Xuan2011} & 8 &17 &90& Plants \\
        \citet{Yousef2015} & - & 114/99 &-& Plants \\
        \citet{Batuwita2009} & 8 &17 &23& Human \\
        \citet{Xue2005} & - &- &32& Human, Plants, Animals, Viruses\\
        \citet{Engchuan2012} & - & 16 & 41 & Plants \\
        \citet{Silla2010} & 2 &17 &10& Plants \\
        \citet{Xuan2011_1} & - &90 &70& Plants \\
        \citet{Gudys2013} & - & 2 & 26 & Human, Plants, Animals, Viruses \\
         \citet{Cui2015} & - &343 &97& Plants \\
    \hline
    \end{tabular}
    }
    \label{tab:featurecount}
\end{table*}

The number of features used in machine learning models is different from model to model (Table \ref{tab:featurecount}). All the features that are clearly mentioned in the literature are listed in Table \ref{tab:featurecount}. 

Table \ref{tab:featurecount} shows that only 9 studies have added thermodynamic related features to their feature set. Shannon entropy, structure entropy, structure enthalpy, and melting energy-related features are the most commonly considered thermodynamical features \citep{Xuan2011,Meng2014,Batuwita2009}. Further, ratios between percentages of each nucleotide over normalized Shannon entropy make another set of features \citep{Williams2012}. Though, thermodynamical features are not much popular among the miRNA studies, these features have entered into the feature lists with the highest information gain \citep{Xuan2011,Meng2014,Williams2012}.

The study \cite{Cui2015} has calculated 440 features under 19 structural and sequence-based feature types. These features include miRNA length, nucleotide frequency, dinucleotide frequency, type of nucleotides surrounding 5 prime arm of the hairpin and 3 prime end, mfe, paired/unpaired nucleotides, presence of bu-lges, base-pairs, triplet elements, and loop related features. As shown in the Table \ref{tab:featurecount}, the studies \cite{Meng2014,Xuan2011_1,Xuan2011} have also used high number (more than 100) of features. Out of 152 features in \cite{Meng2014}, 69 features were novel while the rest are from previous studies like triplet-SVM, miPred, microPred, PlantMiRNAPred. MicroRNA prediction model called PlantMiRNAPred has used 115 selected features under 3 categories; primary sequence-related feature subset, secondary structure related feature subset, and energy \& thermodynamic related feature subset. As shown in Table \ref{tab:featurecount}, the studies \cite{Douglass2016} and \cite{Tseng2018} have used the minimum number of features in their studies. Sequence length, read counts in the sequencing library, the number of mapping locations of the sequence on its respective genome (multiplicity), entropy, and the existence of detectable predicted miRNA* sequences are the features considered in this study. One sequence feature and four structure-related features have been considered to construct the model by the study \cite{Tseng2018}. The sequence feature is the existence of the mature miRNA (5’ - uracil). Read count difference between guiding and passenger strands, mfe, triplet elements, and the dicer cutting site are the four structure-related features. MicroRPM \citep{Tseng2018} has tested different combinations of these five features to selected efficient features for the prediction model. The triplet elements and pairing structure at Dicer cutting sites have been identified as more significant features for miRNA prediction than the read count difference between guiding and passenger strands \citep{Tseng2018}.

Base-pair compositions, mfe related features, triplet elements, structural diversity, base-pair distance, the average number of mismatches per 21-nt window, and structural compactness were identified as the most commonly used structure-related features. All most all the studies considered in this review have used 16 dinucleotide frequencies and G+C\% content as sequence-based features. 

The other notable sequence-based feature is the sequence motif. The sequence motif is a short stretch of nucleotides that is widespread among plant pre-miRNAs. Motif discovery, in turn, is the process of finding such short sequences within a larger pool of sequences; here in plant hairpins. Among the selected studies, only 3 have used the motif sequences in their feature set \citep{Sunkar2008,Yousef2015,Yousef2015a}. The study \cite{Yousef2015} and MotifmiRNAPred \citep{Yousef2015a} have considered precursor sequence motifs as the features for the model. To discover motifs in both positive and negative sequences, a web-server called MEME Suite (Multiple EM for Motif Elicitation) has been used \citep{Bailey2009}. The study \cite{Yousef2015a} shows that for plant miRNA detection, motif-based features along with sequence-based features have lead them to a good recognition of pre-miRNAs. Though several studies have reported that the set of features used in their model is the best for this particular task, none of them have provided pieces of evidence to claim their verdict. 

Some of the studies have not used all the calculated features to construct the ML model. SVM recursive feature elimination (SVM-RFE) is the most popular feature elimination algorithm among the studies which have followed the feature selection step \citep{Yousef2015,Meng2014,Yousef2015a}. The study, \cite{Meng2014} has followed the Back-Support Vector Machine-Recursive Feature Elimination (B-SVM-RFE) method to eliminate redundant features and select informative feature subset. Information gain (IG) has been used in \cite{Wang2019,Gudys2013,Xuan2011,Xuan2011_1,Yousef2016_2} to rank the features. The study \cite{Yousef2016_2} has considered 8 feature selection methods. They are selecting features based on low information gain (LIG), random feature selection (RFS), random feature selection from feature clusters (RFC) feature selection from clusters (SFC), selecting features with high information gain (HIG), the highest information gain selection from feature clusters (HIC), zero-norm feature selection (ZNF) and Pearson correlation-based feature selection (PCF). However, these feature selections were not to identify the impact of the feature selection but to compare the impact of one-class classifiers and two-class classifiers. The studies such as \cite{Jiang2007,Yao2016,Tseng2018} have tested the model with different combinations of features (based on type).

\section{Model validation and evaluation methods}

Model validation is where the trained model is evaluated with the test data. Most of the studies have used cross-validation as the model evaluation technique. 10-fold cross-validation \citep{Vitsios2017,Gudys2013,Yao2016,Guan2011,Engchuan2012} and 5-fold cross-validation  \citep{Batuwita2009,Zhong2015,Koh2017} techniques were identified among the selected studies. The study \cite{Williams2012} have followed leave-one-out cross-validation. The model developed in \cite{Yousef2010} is trained using 90\% of the positive class and the remaining 10\% is used for sensitivity evaluation without using cross-validation techniques. Conducting an experimental method to examine the predicted miRNAs adds value to the studies of novel miRNA identification. However, only three studies have experimentally validated their results \citep{Tseng2018,Wang2004,Sunkar2008}. 

The model evaluation step is an essential part of any machine learning model implementation. Various performance metrics define the behavior of the implemented models. Most of the studies have used accuracy, sensitivity, specificity, Matthews Correlation Coefficient (MCC), geometric mean, and recall rate metrics to measure performances of the computational models \citep{Batuwita2009,Wu2011,Yao2016,Xuan2011,Guan2011,Yousef2010,Jiang2007}. Area Under the Receiver Operating Characteristics (AUC-ROC) curve is taken into account as another performance measurement \citep{Douglass2016,Vitsios2017,Tseng2018}. It indicates the capability of the model to distinguish between classes. MicroRPM has used both validation accuracy and the area under the ROC plot curve to evaluate the model performance.

All the studies have used different datasets to train their models. So it is not straightforward to compare the accuracies of those algorithms. But some of the studies have compared their models with others using a common test data \citep{Guan2011}. Table \ref{tab:performance} is taken from \cite{Guan2011} with some modification. It describes the Accuracy, Sensitivity, and Specificity values of four studies where the testing set is the same but the training set is different.

\medskip

\begin{Table}\centering 
\tcap{Performance of different models. SP:Specificity SE:Sensitivity Acc:Accuracy
A:Triplet-SVM, B:microPred, C:MiPred, D:mirExplorer
(Training datasets used for these studies are different)}
{\small
\setlength{\extrarowheight}{3pt}
\baselineskip=13pt
\begin{tabular}{l c c c c}  \hline
        Model & ML Model & Acc (\%) & SP (\%) & SE (\%) \\
    \hline
        A & SVM & 70.60 & 83.50 & 88.40 \\
        B & SVM & 88.18 & 66.43 & 90.50 \\
        C & RF & 80.98 & 84.34 & 93.56 \\
        D & Adaboost & 92.22 & 97.11 & 94.32 \\
    \hline
\end{tabular}}
\label{tab:performance}
\end{Table}

\section{Discussion}

The performance of the study \cite{Koh2017} is compared with the studies \cite{Jiang2007} and \cite{Batuwita2009} in means of sensitivity and specificity through 5-fold cross-validation. Though all of them use the same negative set, a direct performance comparison cannot be done because they used different positive datasets. Furthermore, the study \cite{Koh2017} has led to high sensitivity and low specificity which can be caused due to the class imbalance in the datasets.

In the study \cite{Yousef2015}, the one-class classifier has achieved higher performance than the two-class classifier wherein both of them have used sequence motif features. Moreover, the one-class classifier trained with motifs has slightly over-performed the SVM model trained with traditional features \citep{Yousef2015}.

The evaluation results of \cite{Tseng2018} indicate that the features such as mfe, triplet element, and dicer cutting site have performed well for both with and without sequence models than the other extracted features. Further, the different combinations of the features have been selected for the model training. Furthermore, the combination of triplet element and dicer cutting site has a good performance for all testing sets without a reference sequence.

\section{Conclusion}

In this review, we focus on the application of machine learning techniques used in plant miRNA identification. With the rise of the machine learning approaches, the possibility to discover novel plant miRNAs accurately became feasible.  Machine Learning approaches have revealed the presence of many non-homologous miRNAs. However, the accuracy of the ML models directly depends on the positive and negative data used in the training process. Also, experimental methods such as northern blotting or reverse transcription PCR are required to validate the results of ML models.  However, most of the studies haven't conducted an experimental validation for their results. 

All of the work discussed in this systematic review have constructed microRNA prediction models using supervised learning algorithms. They have followed different criteria to divide the dataset into positive and negative controls.  However, most of the microRNA in the online repositories for miRNA sequences and annotations, have not been experimentally validated for their existence in the plants. Therefore, dividing the data set as positive and negative controls may affect the performance of the model. Hence, a model based on semi-supervised learning algorithms may give high performance. In the semi-supervised learning approach, a small amount of labeled data is combined with a large amount of unlabeled data during the training. As mentioned before, selecting a negative dataset has a huge impact on model performance. The negative datasets are generated with non-coding RNAs such as tRNAs, rRNAs, etc. But these RNAs do not ensure that they are negative data. Therefore, one of the main drawbacks is the lack of criteria to select confident negative training sets, and the negative sequence dataset generation method will affect the prediction results. Therefore, a more precise technique to generate a negative dataset is required.

When selecting the studies for this review, we found that there are very few studies using a machine learning approach (non-comparative) with only plant data. This shows the need for plant-specific or species-specific computational methods in miRNA identification. Compared to the datasets available for animal species, humans and viruses, there is a relatively small number of plant miRNAs that have been experimentally verified. Therefore, we encourage computer science, data science and biology experts to work together since computational tools are tightly linked to experimental biological research. Improved understanding of molecular mechanisms of miRNA in plants will lead to developing novel and more precise techniques.

The audience who are benefited from microRNA prediction models are non-experts in the computer science or data science field. Therefore, it would be effective and useful if the developed models are integrated with user-friendly software. Having that, the intended audience can get the benefits of the developed models by using them for experimental work and it will add more value to the proposed workflows as well. 

\section*{Author contribution statement}

PI, SL, and SJ outlined and collected the literature associated with the review, performed evaluations of the literature, and drafted the manuscript. DH supervised the work and contributed to the preparation of the manuscript. All authors read and approved the final manuscript.

\section*{Conflict of interest}

The authors declare that they have no conflict of interest.

\section*{Acknowledgments}

The authors would like to thank Asst. Prof. Indika Kahanda, School of Computing, University of North Florida, USA and Mrs. Buwani Manuweera, PhD Student, Montana State University, USA for providing assistance and guidance in preparing the manuscript.


\bibliographystyle{unsrtnat}
\bibliography{ref}

\end{multicols}
\end{document}